\documentclass{article}

\usepackage{microtype}
\usepackage{graphicx}
\usepackage{subfigure}
\usepackage{booktabs} %

\usepackage{hyperref}

\usepackage{mlsys2025}

\usepackage{listings} %
\usepackage{amsmath}  %
\usepackage{xspace}
\usepackage{hyperref}
\usepackage{xcolor,cleveref}
\usepackage{graphicx}  %
\usepackage{pifont}  %
\usepackage{marginnote}
\usepackage{algorithm}
\usepackage{algpseudocode}
\usepackage{multirow}
\usepackage{tcolorbox}
\usepackage{enumitem}
\usepackage{amsthm}
\usepackage{tcolorbox}
\usepackage{makecell}

\newtcolorbox[auto counter]{protocol}[2][]{
  colback=white, colframe=black,
  fonttitle=\bfseries,
  sharp corners,
  title=Protocol~\thetcbcounter: #2, #1
}

\definecolor{dkgreen}{rgb}{0,0.6,0}
\definecolor{gray}{rgb}{0.5,0.5,0.5}
\definecolor{mauve}{rgb}{0.58,0,0.82}

\crefalias{formula}{equation}
\crefname{formula}{formula}{formulas}
\Crefname{formula}{Formula}{Formulas}

\newif\ifsubmit
\submitfalse

\newcommand{\sysname}{DeServe\xspace}

\newcommand{\remove}[1]{}

\newcommand{\reviewer}[3]{
  \expandafter\newcommand\csname #1\endcsname[1]{
    \ifsubmit
      \ignorespaces
    \else
      \textcolor{#3}{[#2: ##1]}
    \fi
  }
}

\reviewer{nini}{Ni}{purple}
\reviewer{brafalsk}{brafalsk}{purple}
\reviewer{xiaoyuan}{Xiaoyuan}{orange}
\reviewer{karen}{Karen}{blue}
\reviewer{tianneng}{Tianneng}{violet}
\reviewer{zhe}{Zhe}{teal}

\mlsystitlerunning{\sysname: Towards Affordable Offline LLM Inference via Decentralization
}

\begin{document}

\twocolumn[
\mlsystitle{\sysname: Towards Affordable Offline LLM Inference via Decentralization
}

\mlsyssetsymbol{equal}{*}

\begin{mlsysauthorlist}
\mlsysauthor{Linyu Wu}{equal,ucb}
\mlsysauthor{Xiaoyuan Liu}{equal,ucb}
\mlsysauthor{Tianneng Shi}{ucb}
\mlsysauthor{Zhe Ye}{ucb}
\mlsysauthor{Dawn Song}{ucb}
\end{mlsysauthorlist}

\mlsysaffiliation{ucb}{University of California, Berkeley}

\mlsyskeywords{Machine Learning, MLSys}

\vskip 0.3in

\begin{abstract}
The rapid growth of generative AI and its integration into everyday workflows have significantly increased the demand for large language model (LLM) inference services. While proprietary models remain popular, recent advancements in open-source LLMs have positioned them as strong contenders. However, deploying these models is often constrained by the high costs and limited availability of GPU resources.
In response, this paper presents the design of a decentralized offline serving system for LLM inference. Utilizing idle GPU resources, our proposed system, \sysname, decentralizes access to LLMs at a lower cost. \sysname specifically addresses key challenges in optimizing serving throughput in high-latency network environments. Experiments demonstrate that \sysname achieves a 6.7x-12.6x improvement in throughput over existing serving system baselines in such conditions.

\end{abstract}
]

\printAffiliationsAndNotice{\mlsysEqualContribution} %

\section{Introduction}

Interest in generative AI, especially text-generating chatbots, has increased significantly recently. Millions of individuals~\cite{yougov} and many companies~\cite{companyUseGPT} are now using AI chatbots in their daily workflows, boosting the demand for large language model (LLM) services.
Closed-source models are popular, but open-source models have recently improved and now show competitive performance in chatbot evaluations~\cite{chiang2024chatbot}. This has drawn industry attention, leading to the emergence of companies providing paid services for open-source LLM inference.

However, despite the wide applicability of LLM services, the platform inference price is typically still too high for daily usage.
Users who want to self-host open-source models to reduce costs have two options to consider: personal devices or rented cloud machines.
Personal devices, even with powerful consumer-grade GPUs, often cannot locally run large models due to memory constraints and can only efficiently run smaller or compressed models with lower performance.
On the other hand, cloud GPU machines are also very expensive for long-term usage.

\begin{figure*}[h]
    \centering
    \includegraphics[width=0.8\textwidth]{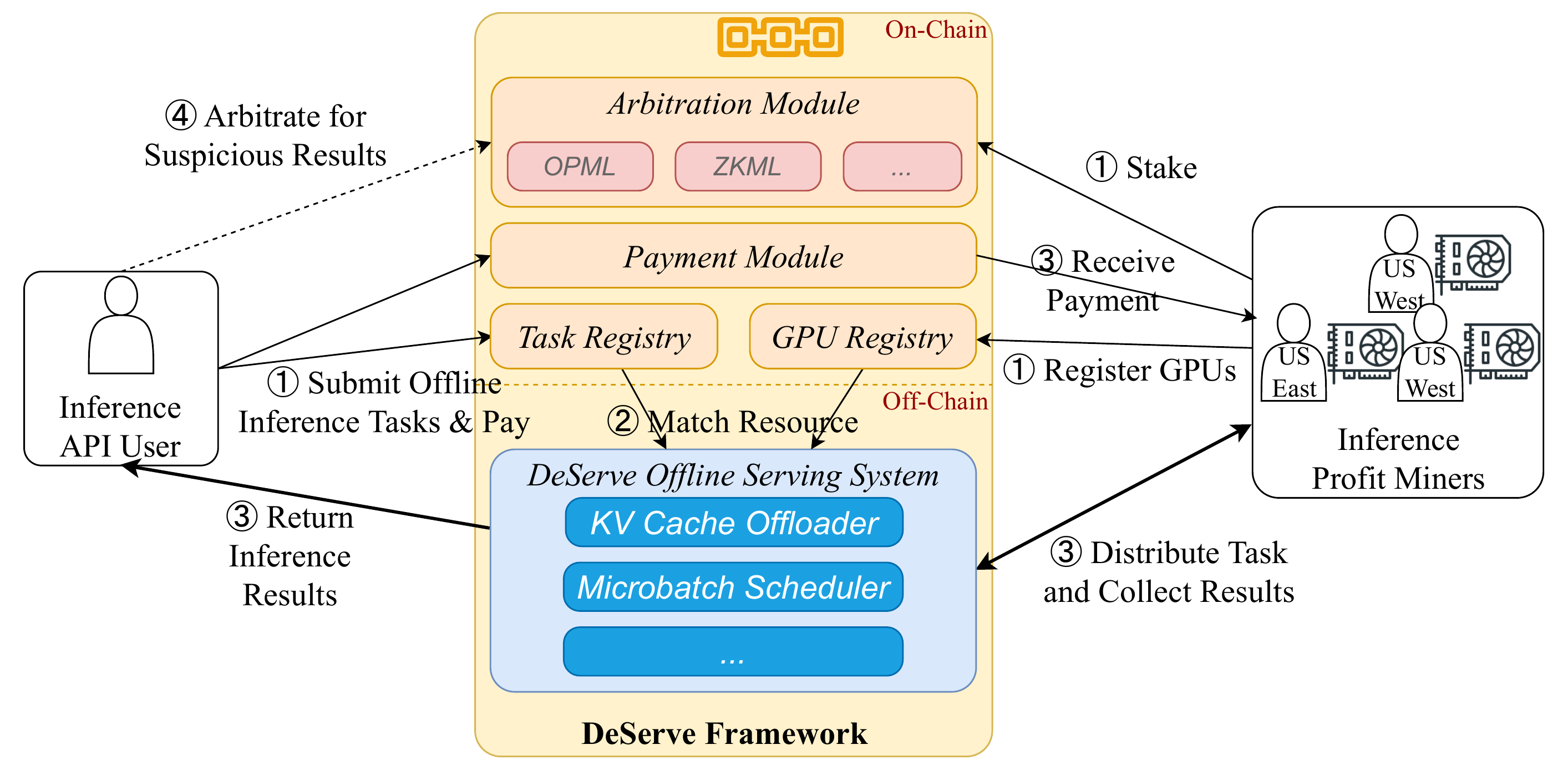}
    \caption{\sysname Framework Overview. It matches users with LLM inference needs and miners with GPUs, while facilitating payments and inference correctness arbitrations.}
    \label{fig:arch}
\end{figure*}

Instead of self-hosting the model, a potential solution is for multiple personal device owners to collaboratively host the model together. By connecting GPUs in a decentralized network, the group may have enough GPU memory to run LLM inference tasks much more efficiently than running them separately. Such a decentralized computation paradigm has also been explored in prime number search~\cite{GIMPS} and crypto mining~\cite{nakamoto2008bitcoin}.
Furthermore, by integrating with decentralized on-chain payment, it is possible to build a framework that matches users who have inference needs with miners who possess compute resources. This concept is illustrated in Figure \ref{fig:arch} and detailed in Section \ref{sec:sec:frame}.

However, the high network latency in decentralized environments prevents existing serving systems from achieving high throughput. To adapt to network conditions and reach practical performance, in this work, we have built a serving system named \sysname that tackles the throughput optimization problem in a decentralized environment. Specifically, our contributions are threefold.

\begin{enumerate}
    \item We identify major performance challenges caused by high-latency decentralized environments and design an efficient offline serving algorithm suitable for such environments. To the best of our knowledge, this is the first work that optimizes the throughput of the offline serving system in a high-latency network environment.
    \item We conduct both simulated and real-world experiments to validate our algorithm design. Our results demonstrate that, compared to the existing serving system baseline, the \sysname system achieves a 6.7x to 12.6x throughput improvement in high-latency environments.
    \item We discuss the correctness protection problem in decentralized collaboration and design a modular \sysname framework for practical deployment integrating with on-chain components.
\end{enumerate}

\section{Background}
\label{sec:bg}

This section covers the basics of LLM serving and model pipelining, with a focus on the network architecture of the widely-used Llama 3 model~\cite{dubey2024llama} as an example.

\subsection{Transformer Layers and Auto-regressive Decoding}
As a text generation model, the Llama 3 model takes the input of text and continuously generates sequences of tokens in a pre-determined dictionary. The model follows an auto-regressive decoding procedure~\cite{graves2013generating}, generating one token in each round of inference until reaching a special termination token (eos). In each round, the previously generated tokens are appended to the input. During the inference, the input is processed through different network layers to generate the probabilistic distribution on the dictionary for the next token. Then, a sampling procedure decides which token should be chosen next. To optimize performance, the model employs a KV cache~\cite{pope2023efficiently}, which stores intermediate states from previous tokens, to reduce redundant calculations and speed up the inference procedure. %

The Llama 3 model consists of three types of layers: one embedding layer, multiple transformer layers, and one output layer\footnote{The last normalization operation is included.}. The number of transformer layers varies depending on the size of the model. In practice, the processing of these transformer layers takes most of the time during the inference procedure. To run model inference efficiently, the model parameters for these layers as well as the KV cache are often loaded into GPU memory when a serving system starts. However, it's important to note that the KV cache can consume a significant amount of GPU memory, especially for long sequences. For instance, in the Llama 3 70B model, the KV cache for a sequence of length 4096 can occupy 1.25 GB of GPU memory. The large GPU memory consumption potentially limits the number of concurrent requests that can be processed simultaneously, which is commonly referred to as the batch size. %

\subsection{Online and Offline Serving System}

An LLM serving system usually takes some inference requests as input and returns responses as output, aiming to optimize the performance. %
Online and offline serving are two different approaches for delivering inference results, each optimized for different operational needs. 
Online serving, also known as real-time inference, is designed for scenarios that need quick responses like chatbots or interactive systems where users expect quick feedback. It often optimizes for low latency, allowing users to interact with the LLM in real time and receive prompt responses. %

On the other hand, offline serving, also called batch inference, is used when processing large volumes of data without the need for instant results like bulk text analysis or generating responses for a large dataset. Its optimization methods prioritize throughput and efficient use of computational resources. %
It's often more cost-effective for processing substantial amounts of data. An example of the cost benefits of offline serving is OpenAI's batch API, which offers a 50\% discount compared to their standard API pricing for users willing to wait up to 24 hours for responses~\cite{openaibatch}. The evaluation metrics for such an offline serving system usually target total throughput.

In a decentralized environment, achieving competitive low latency compared to a centralized serving system is challenging due to inherent network latency issues. A more practical approach is to focus on the offline serving setting. By orchestrating the inference pipeline across decentralized nodes and leveraging idle periods during the network transfer, the system can mitigate the impact of high network latency and improve the total throughput.

\subsection{Model Parallelism and Partition}
\label{sec:bg:parallelism}

Limitation in GPU memory is the major challenge when serving a large model such as the Llama 3 70B, which needs approximately 130 GB of GPU memory. However, widely-used GPUs, like H100, only have less than 80 GB of GPU memory, not to mention that consumer-grade GPUs have even less memory. As a result, it is often infeasible to serve the entire model using a single device. 

To address this challenge, the widely adopted approach is model parallelism, a technique commonly employed in both 
distributed training~\cite{shazeer2018mesh,shoeybi2019megatron,zheng2022alpa},
and serving~\cite{yu2022orca,li2023alpaserve} phases. 
For transformer-based LLMs, there exist two primary categories of model parallelism, differentiated by their methods of partitioning the models: intra-layer and inter-layer parallelism. We introduce them below and discuss why we choose inter-layer parallelism in the decentralized environment.

\paragraph{Intra-layer Parallelism}
Transformer models perform high-dimensional matrix multiplications on tensors. Intra-layer parallelism, also known as tensor parallelism, splits these operations and their parameters over multiple devices~\cite{shoeybi2019megatron}.
This approach allows faster computation on large models that wouldn't fit on a single GPU, making it particularly suitable for models with massive parameter sizes. 

\paragraph{Inter-layer Parallelism}

Since a transformer model is executed layer by layer, the inter-layer parallelism divides the model into smaller stages with a layer as the smallest unit and then distributes them to different devices~\cite{zheng2022alpa}.
The stages are executed in a pipeline way~\cite{huang2019gpipe,li2021terapipe} to avoid waiting for the completion of the previous stage and reduce the idle time of devices during multiple rounds of execution. So it can also be referred to as pipeline parallelism. This strategy substantially reduces communication overhead because only a small amount of data needs to be forwarded from the output of one stage to the input of the next stage. 

Considering that intra-layer parallelism can lead to large communication overhead due to data exchange between GPUs~\cite{li2023alpaserve}, a decentralized serving system can derive significant benefits from inter-layer parallelism, especially under slow and unreliable network conditions. %
Thus, when serving models that require GPU memory exceeding the maximum available GPU memory in any node in the network, we select a proper set of nodes and use model pipelining to distribute the layers.
We will then dive into finer-grain optimizations in \Cref{sec:opt}.

\section{Motivation and Cost Model}

A LLM-serving system processes users’ inference tasks, generating tokens for each task until reaching the special termination token, thereby earning the corresponding user payment. As an offline-mode service, given a set number of tasks and their associated profit, the serving system consumes hardware compute power within a specific time frame as the cost in exchange for revenue. The construction of a serving system is well motivated when hardware costs remain below revenue.

In this section, we first separately explain the cost of different compute resources and the LLM service pricing model. Then, combined together, we calculate the profit requirement for a LLM serving system.
Based on the profit requirement, we motivate the need for decentralized serving.

\subsection{Cost of Different Compute Resources}
Running large models,
such as llama-70b which consumes large amounts of GPU memory, often requires multiple GPUs across several machines. Such resource cost estimation generally follows two approaches. 

The first approach considers direct procurement, listing corresponding hardware purchase prices, space requirements, power costs, and estimated lifespan to calculate comparative compute power. This approach is centralized and benefits from relatively stable costs. However, this long-term estimation approach is less feasible given the dynamic compute demands of continuously evolving models and the increasing power of ever-releasing GPUs.

The second approach is to consider the dynamic pricing of pay-per-use resources, such as renting GPU instances on cloud platforms. This approach outsources the main cost components to the platform, allowing for relatively flexible price comparisons.
However, due to different platform profit strategies, costs can vary significantly depending on the compute resource provider. This cost model is more practical for profit reasoning. 
Thus, in this work, we mainly focus on the second approach. 

Specifically, we consider three distinct sources of compute resources: cloud platforms, decentralized compute platforms, and the decentralized mining paradigm.

\begin{enumerate}
    \item \textit{Cloud platforms} (e.g., AWS~\cite{aws}, Azure~\cite{azure}, GCP~\cite{gcp}) typically provide GPU resources within data centers. Their advantages include publicly accessible, relatively stable services with good network conditions, high internal bandwidth, and low latency. However, they generally offer a limited range of GPU types and are relatively expensive, sometimes with rental quotas. 
    Overall, these resources are more suitable for low-latency online serving, which values stability and good network conditions.
    \item With the rise of the decentralized ecosystem, new \textit{decentralized compute platforms} (e.g., RunPod~\cite{runpod}, io.net~\cite{io.net}) allow individuals to contribute compute power to the platform, offering pay-per-use resources similar to cloud platforms. Compared to cloud platforms, decentralized platforms typically feature lower prices and a wider selection of GPUs. However, the contributed hardware tends to vary in quality, and the platform often ensures computational isolation via containerization and may employ firewalls or specialized traffic routing mechanisms. As a result, there have been operational issues with stability (e.g., instances frequently going offline unexpectedly) and internal networking (e.g., lack of interconnectivity among machines). %
    \item The \textit{decentralized mining paradigm} manages compute resources by distributing specific executable binaries that connect with user devices to execute dedicated protocols. This paradigm has been widely applied, from voluntary prime number searches (e.g., GIMPS~\cite{GIMPS}) to profitable cryptocurrency mining (e.g., Bitcoin~\cite{nakamoto2008bitcoin}). Today, GPU-based mining projects continue to attract users with available GPU hours, offering opportunities to earn revenue, with some websites (e.g., What to Mine~\cite{whattomine}) rank these projects by profitability. From another perspective, this highlights users’ willingness to contribute compute power at mining rates. Compared to the previous two sources, mining-based estimates yield lower costs. However, promoting usage and managing the network introduce extra challenges.
\end{enumerate}

Summarizing the discussion above, Table~\ref{tab:motivation_quan} presents a qualitative comparison of the three sources of compute resources. Focusing on offline serving, in this work we mainly focus on estimations based on the cost of the latter two sources. 

\begin{table*}[th]
    \centering
    \begin{tabular}{c|c|c|c}
        \hline
        Compute Platform & Cloud Platform & Decentralized Compute Platform & Decentralized Mining \\ \hline
        Network Latency & Low & Medium & High \\  \hline
        GPU Type & Standardized & Heterogeneous & Heterogeneous \\ \hline
        Service Availability & 99.9\%{}+ Uptime & Variable Uptime & Intermittent \\ \hline
        Compute Cost & Expensive & Cheaper& Cheapest \\
        \hline
    \end{tabular}
    \caption{Comparison of key characteristics across different computing platforms. }
    \label{tab:motivation_quan}
\end{table*}

\subsection{Pricing for LLM Services}

Users typically access LLM services via a web UI or API, with the former involving subscription-based payment and cost estimates based on usage frequency. Therefore, we primarily focus on API services here.
There are two major pricing models for LLM API services: on-demand pricing and provisioned pricing. The former charges based on token usage (e.g., OpenAI~\cite{openaiprice}, Anthropic~\cite{anthropicprice}), while the latter provides a stable maximum throughput for a committed period (e.g., AWS Bedrock’s provisioned model~\cite{bedrockprice}). We primarily consider on-demand pricing, given that most platforms adopt this model.

The on-demand pricing model varies based on the service scenario and the modeling of computational tasks. Service scenarios include online interfaces for immediate returns and asynchronous offline-serving interfaces. For example, OpenAI’s batch API allows up to 24 hours to return results at half the price of its online version~\cite{openaibatch}. In terms of task modeling, due to differences in resource utilization between the prefill stage (input processing) and the decode stage (output generation)~\cite{agrawal_taming_2024}, some platforms differentiate pricing between input tokens and output tokens, while others use a unified pricing strategy for simplicity. 
Next, we adopt this simplified model to discuss the profit requirement for a serving system.

\subsection{Profit Model}
With an understanding of the cost model and existing pricing structures, we now construct the profit model for an offline LLM serving system. The key is to parameterize the conversion rate from compute resources to token processing, reflecting the serving system's performance.

Consider a single compute resource unit forming a pipeline that serves the target LLM, with a per-time cost $C$ from a specified source. For a given workload of requests containing $N_I$ input tokens and expecting $N_O$ output tokens, if the system completes the workload with $N = N_O + N_I$ tokens in total within time $T$, we evaluate its performance using two metrics: input throughput, $M_I = \frac{N_I}{T}$, and output throughput, $M_O = \frac{N_O}{T}$. To align with a simplified pricing model using unified input and output pricing, we consider the total throughput $M = M_I + M_O$.

Assuming the workload follows an on-demand pricing model with input price per token $P_I$ and output price per token $P_O$, the total revenue generated is $R = N_I \cdot P_I + N_O \cdot P_O$. Alternatively, under a simplified pricing model where $P = P_I = P_O$, we have $R = N \cdot P$.

The system is profitable when $R > C \cdot T$, which sets a performance challenge. For example, in the simplified pricing model, the system must satisfy the following requirement: $R > C \cdot T \Leftrightarrow M > \frac{C}{P}$.
Table~\ref{tab:motivation_example} displays the calculated profit thresholds across various settings.
As shown in Table~\ref{tab:motivation_example}, mining platforms offer dramatically lower compute costs at just \${}0.35 per hour compared to \${}13.88 for cloud platforms, requiring only 108 tokens/second throughput to break even - making them an attractive option for cost-effective LLM inference if throughput and integrity challenges can be addressed.

\begin{table*}[th]
    \centering
    \begin{tabular}{c|c|c|c|c}
        \hline
        Compute Platform & Cloud Platform & \multicolumn{2}{c|}{\makecell{Decentralized\\Compute Platform}} & \makecell{Decentralized Mining} \\ \hline
        Compute Spec & \makecell{GCP-8x g2-standard-32\\(with L4)} & RunPod-8x4090 & io.net-8x4090 & WhatToMine-8x4090 \\  \hline
        \makecell{Compute Cost\\($C$, \${} per hour)} & 13.88 & 5.52 & 3.69 & 0.35 \\ \hline
        \makecell{Example Inference Price\\($P$, \${} per 1M token)} & \multicolumn{4}{c}{0.90} \\ \hline
        \makecell{Throughput Requirement\\($M_{\text{min}} = \frac{C}{P}$, token per second)} & 4,283.33 & 1,703.70 & 1,138.89 & 108.02 \\
        \hline
    \end{tabular}
    \caption{Cost analysis across different compute platforms with equivalent GPU count (8x). We use NVIDIA L4 for GCP due to lack of 4090 availability. The example inference price is from Together.ai's Llama-70B model. The minimum throughput requirement shows how many tokens per second each platform must process to profit.}
    \label{tab:motivation_example}
\end{table*}

Since decentralized mining provides the lowest compute cost, there is a strong incentive to build a serving system that achieves high throughput under this paradigm. 
However, existing serving systems often suffer from significant throughput degradation in such environment. Next, in Section \ref{sec:opt}, we discuss our major techniques to improve throughput in the decentralized environment.

\section{Optimize Decentralized Serving}
\label{sec:opt}

Compared to centralized serving, decentralized environments often introduce higher network latency and limited bandwidth.  
For instance, for GPUs located in US east and US west to collaborate, it takes around 60 milliseconds for a packet to arrive from one to the other. 
The high network latency between GPUs greatly impacts inference efficiency. For example, using pipelining in vLLM between machines across US, there is a 58\% reduction in overall throughput compared with centralized setting.

Additionally, decentralized settings often involves more consumer-grade GPUs with less memory. Thus, serving a model with the same memory requirement takes more GPUs.
For instance, an NVIDIA H100 has 80GB of memory, while the RTX 4090 offers only 24GB. As a result, two H100 GPUs can serve a 70B model, whereas it requires eight RTX 4090 GPUs to handle the same workload.
These factors complicate efforts to maintain high throughput in decentralized serving. 
Specifically, we identify two major challenges for the serving algorithm design.

\paragraph{Limited memory for each batch.}
In pipeline parallelism, multiple microbatches are present on each machine, but only one is actively executed at any given time. 
Since these microbatches share the same GPU memory, each can only own a portion of it. 
Consequently, the batch size for each microbatch is limited, which in turn restricts the overall GPU utilization rate. 
In Table~\ref{tab:only_decode}, we illustrate the relationship between batch size and execution time for decoding with same prefix.

\begin{table}[h]
    \centering
    \begin{tabular}{c|c|c}  %
        \hline
        Batch Size & Total Time (ms) & Time Per Instance (ms) \\  %
        \hline
        1 & 66.6 & 66.6 \\
        \hline
        2&	68.9&	34.5 \\
        \hline
        4&	69.1&	17.2 \\
        \hline
        8&	69.5&	8.69 \\
        \hline
        16&	70.3	&4.39 \\
        \hline
        32	&76.5&	2.39\\
        \hline
        64&	80.2 &	1.25\\
        \hline
        128&	89.1&	0.696\\
        \hline
        256	&137.5&	0.537\\
        \hline
    \end{tabular}
    \caption{Execution time with different batch size.}
    \label{tab:only_decode}  %
\end{table}

\begin{figure*}[th]
    \centering
    \includegraphics[width=0.8\textwidth]{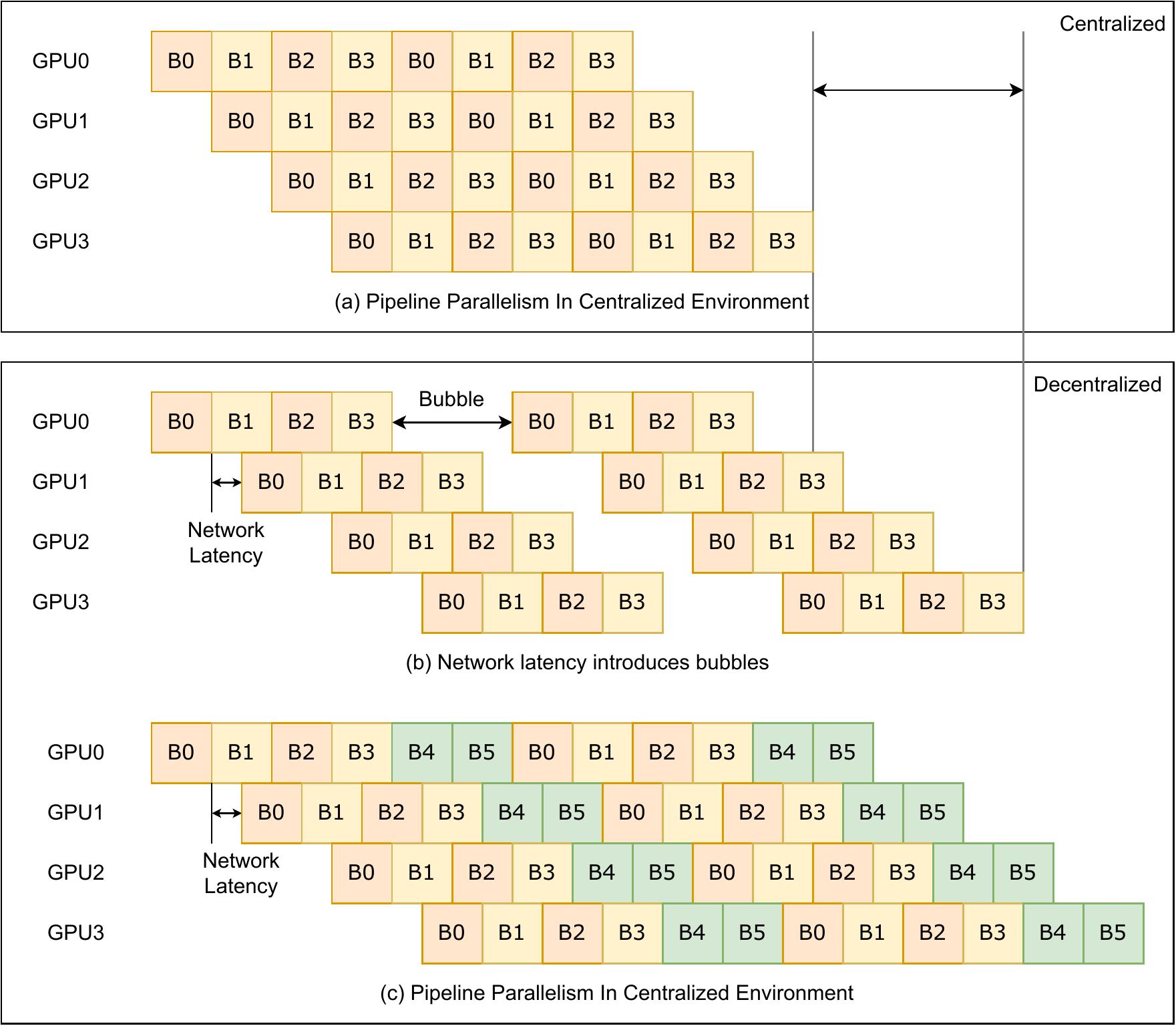}
    \caption{Pipeline parallelism in centralized and decentralized environment.}
    \label{fig:pipeline}
\end{figure*}

\paragraph{Network latency bubbles.} 
High-latency networks have a significant impact on the efficiency of LLM inference. 
As illustrated in Figure~\ref{fig:pipeline}(a), there is no idle periods, i.e. \textit{bubbles}, in centralized environment, microbatches are executed sequentially and the GPUs are kept highly utilized. 
However, as illustrated in Figure~\ref{fig:pipeline}(b), the network latency introduces bubbles inside the pipeline, wasting available GPU computation power. Bubbles greatly impact the GPU utilization rate, leading poor efficiency of LLM inference.

\subsection{Overview}

We highlight two major designs in \sysname to overcome the above challenges: KV cache offloading and microbatch scheduling.

KV cache offloading increases available GPU memory by transferring unused KV cache to CPU memory, enabling larger batch sizes per microbatch and more efficient memory allocation. Additionally, the system adjusts the number of microbatches in the pipeline to fill bubbles caused by network latency, ensuring optimal GPU resource utilization.

\subsection{KV Cache Offloading}
\label{sec:opt:offloading}

\begin{figure}[th]
    \centering
    \includegraphics[width=0.35\textwidth]{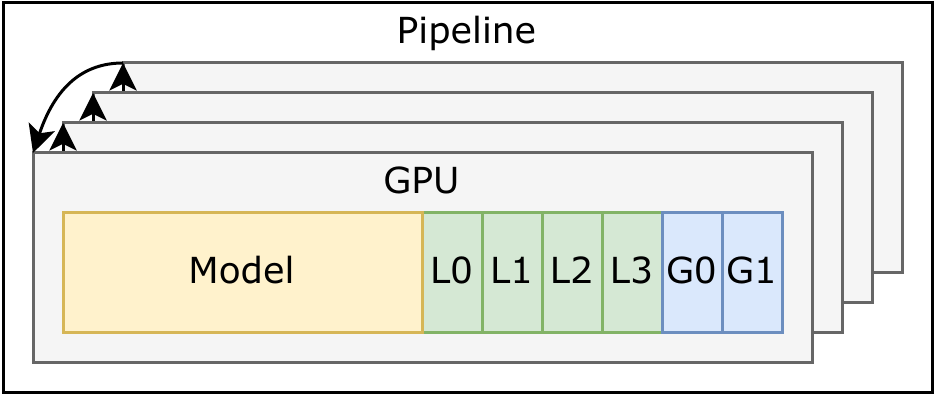}
    \caption{Memory layout of DeServe with 4 microbatches. Model represents the model weights. L0, L1, ... represent local page pools for each microbatch. G0, G1 represent global page pools.}
    \label{fig:mem-layout}
\end{figure}

We illustrate the GPU memory layout during serving in Figure~\ref{fig:mem-layout}. The memory layout accommodates model weights and KV cache similar to some existing works~\cite{strati_dejavu_2024}. 
Specifically, when using $N_M$ machines with $M$ memory and $N_B$ microbatches to serve a Llama 3 70B model, the distributed model weights occupy $M_W$, leaving $M_{KV} = M - M_W$ for KV cache allocation.
Microbatches need to allocate their own KV cache, thus sharing only a portion of GPU memory for KV cache, which is formulated as $M_B = \frac{M_{KV}}{N_B}$.

To increase GPU memory for each microbatch, we use KV cache offloading. Specifically, we define two global page pools, $G_0$ and $G_1$, each with $M_G$ GPU memory, to swap KV cache between CPU and GPU. Additionally, $N_B$ local page pools are allocated for each microbatch, which remain on the GPU without offloading. By using this strategy, each microbatch can utilize one global page pool and its own local page pool. Thus, available GPU memory for each microbatch will become:
\begin{align} \label{eq:micro_batch_mem_size}
M_B' = \frac{M_{KV} - 2M_G}{N_B} + M_G
\end{align}

The size of $M_G$ is determined by the PCIe bandwidth $W$ and pipeline stage time $T_S$.
\begin{align} \label{eq:global_page_pool_size}
    M_G = W \times T_S
\end{align}

\begin{figure}[th]
    \centering
    \includegraphics[width=0.45\textwidth]{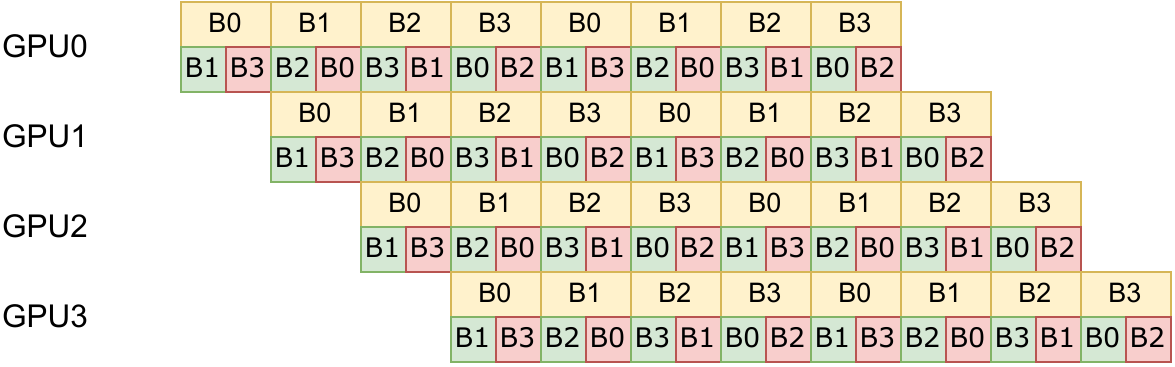}
    \vspace{-0.4cm}
    \caption{Overlapping KV cache swapping with LLM inference. LLM inference computation is represented in yellow, KV cache swapping in is in green, and KV cache swapping out is in red. }
    \label{fig:deserve-swap}
\end{figure}

To avoid overhead for KV cache offloading, we overlap KV cache offloading with LLM inference computation. For example, when microbatch $B_1$ is being executed with its local page pool and global page pool $G_1$ for KV cache, we simultaneously offload previous KV cache used by $B_0$ and prefetch KV cache for $B_2$ in $G_0$ utilizing PCIe's full-duplex capability, as illustrated in Figure~\ref{fig:deserve-swap}.

By leveraging KV cache offloading, larger batch size is available for each microbatch, improving GPU utilization and overall system throughput.

\subsection{Microbatch Scheduling}

We propose microbatch scheduling to address pipeline bubbles caused by network latency. This solution is illustrated in Figure~\ref{fig:pipeline}(c). In the example showed in the figure, there are four machines inside the pipeline, and the network latency is $\frac{1}{2} T_S$. The bubbles brought by the network latency can be resolved through adding two more extra microbatches to fill the blank.

By combining microbatch scheduling with KV cache offloading, the system is able to yield larger optimization space. 
Without KV cache offloading, doubling the number of microbatches would halve the average batch size per microbatch, rendering this optimization ineffective. 
However, as shown in Formula~\ref{eq:micro_batch_mem_size}, increasing the number of microbatches still guarantees a lower bound of $M_G$ GPU memory per microbatch, determined by PCIe bandwidth. This ensures a minimum throughput level, regardless of the number of microbatches we use to cater network condition. 

In conclusion, by scheduling microbatches according to the network latency, we can maintain high throughput even in extreme network condition. We provide the experiment result in the next section.

\section{Experiments}

\begin{table*}[h]
\centering
\begin{tabular}{l|rr|rrrr}
\toprule
\multirow{3}{*}{Latency (ms)} & \multicolumn{2}{c|}{real-world} & \multicolumn{4}{c}{simulation} \\ \cline{2-7}
& \multicolumn{1}{c}{centralized} & \multicolumn{1}{c|}{east-west} \\
& \multicolumn{1}{c}{\textless{}1} & \multicolumn{1}{c|}{58.4} & \multicolumn{1}{c}{16} & \multicolumn{1}{c}{32} & \multicolumn{1}{c}{64} & \multicolumn{1}{c}{256} \\ \hline
vLLM (tp)    & 253.0 & failed & / & / & / & / \\ \hline
vLLM (pp)    & 89.1 & 37.3 & 68.8 & 55.3  & 36.1 & /    \\ \hline
\sysname (pp) & 194.6 & 138.4 & 182.3 & 163.7 & 133.7 & /    \\ \hline
\sysname (opt) & 445.2 & 434.1 & 458.5 & 457.3 & 456.8 & 442.9 \\
\bottomrule
\end{tabular}
\caption{Output throughput under different latencies. The \sysname (opt) row shows the effect of optimizations mentioned in Section \ref{sec:opt}.} \label{tab:exp1}
\end{table*}

We use a static workload with an average prompt length of 256, randomly selecting lengths from $[0, 512]$, with generated lengths chosen similarly. We send $N_R$ requests simultaneously for prefilling and decoding, replenishing them as the previous requests are completed. The system is benchmarked for 20 minutes with statistics collected from the last 16 minutes. Since the average input and output lengths are the same, we expect the output throughput to match the input throughput. \sysname is implemented in PyTorch~\cite{torch} with CUDA kernels from FlashInfer~\cite{flashinfer2024}. The codebase for the experiments has been published in \href{https://github.com/CoLearn-Dev/deserve}{https://github.com/CoLearn-Dev/deserve}.

We run real-world experiments by machines across us-east-1 and us-west-4 in Google Cloud with 58.4ms latency, and use the machines in the same region with code patches to simulate latency for other experiments.

Table~\ref{tab:exp1} shows the main experiment result.
The vLLM threw out an error during the benchmark when we tried tensor parallelism in machines across two regions. It might be because of the high latency or insufficient bandwidth between the regions. Thus, pipeline parallelism would be a better choice in a decentralized environment.
As we can see in Table~\ref{tab:exp1}, the east-west and simulated 64ms columns, the real-world experiments with similar simulated latency yield similar results, validating our latency simulation approach.

Comparing our framework with the vLLM pipeline parallelism method, we can see our implementation consistently has better performance than vLLM.
As we observe, the performance drops quickly with increasing latency if we don't apply the KV cache offloading and microbatch splitting described in Section~\ref{sec:opt}. By implementing these optimizations, we can sustain a satisfactory level of performance even as latency increases much higher.

\section{Discussion}

Unlike cloud or decentralized computation platforms, in a decentralized mining paradigm, individual compute resource contributors may deviate from the protocol, providing incorrect values during execution. For instance, malicious participants may falsely claim GPU usage but instead run the protocol on a low-cost CPU instance, producing outputs with the correct protocol format but containing dummy values. Even when using a legitimate GPU instance, these participants could still send incorrect values to alter the output, undermining the service's trustworthiness.

Formally, consider the protocol between a compute resource contributor and an LLM service requester deploying the serving system. An ideal system should ensure the correctness of inference results received by the service requester, such that these results are indistinguishable from those produced by an honest resource provider.

Extensive research has explored methods to guarantee computational correctness among decentralized contributors. In this section, we first examine existing work and summarize key design principles. We then discuss the modular \sysname system design, which integrates these principles to safeguard inference correctness and fairly manage incentives.

\subsection{Protect Inference Correctness}
\label{sec:sec:exist}

Existing protection mechanisms primarily focus on three main approaches: optimistic verification, voting-based consensus, and zero-knowledge proof.
opML~\cite{conway2024opml} employs an optimistic approach to protect LLM inference correctness on the blockchain. The approach is similar to Agatha~\cite{zheng2021agatha} which protects the inference correctness of DNN. It adapts the bisection protocol, similar to Arbitrum's~\cite{kalodner2018arbitrum} method for smart contracts, to pinpoint errors in the computation process, either at the granularity of tensor operations or element-wise operations. 
On the other hand, spML~\cite{zhang2024proof} utilizes a voting-based consensus mechanism. It relies on a decentralized network of validators to determine the accuracy of inference results through a voting process. Both methods need deterministic computation, such as software-based floating-point operations, to ensure consistent results across different hardware platforms.
zkLLM~\cite{sun2024zkllm} introduces a parallelized lookup argument to bring zero-knowledge proof to LLM inference. However, the computation cost still very high.

Additionally, we further identify the following design principles that are vital for the practicability of a decentralized serving system.

\begin{enumerate}
    \item \textbf{The correctness protection should not introduce heavy extra computation costs for the inference service provider.} Both opML and spML rely on the deterministic computation which slow downs the inference, and zkLLM rely on computationally expensive zero-knowledge proofs. All these previous solutions fail to meet this principle.
    \item \textbf{Both inference input and output should remain off-chain by default to avoid extra costs.} Storing large data on-chain often incurs substantial blockchain computation fees (i.e., gas fees), potentially exceeding the cost of the inference itself. This makes solutions like opML, which require storing all results on-chain, cost-prohibitive.
    \item \textbf{The inference service provider should not be required to respond to verification challenges from arbitrary parties.} opML requires the inference service provider to respond to challenges from any verifier, potentially exposing the provider to denial-of-service (DoS) attacks.
\end{enumerate}

\subsection{\sysname Framework}
\label{sec:sec:frame}

Our \sysname framework (cf. \Cref{fig:arch}) is designed with theabove principles in mind.

Based on the above principles, we provide the \sysname framework design in Figure \ref{fig:arch}. 
In addition to the serving system discussed in Section \ref{sec:opt}, the framework introduces four more on-chain components to handle revenue distribution and correctness protection.
The \textit{task registry} and \textit{GPU registry} enable service discovery and record offline inference tasks from the user and available compute resources from miners. 
The \textit{payment module} accepts and locks user payment in cryptocurrencies on-chain and transfers them to the corresponding miner after the inference results are returned to the user.
The \textit{arbitration module} is not triggered during an honest collaboration. However, if the user receives incorrect inference results, given the signature from the miner that produces the result, the user can invoke arbitration procedure that eventually claims the staking of the corresponding miner. 
The requirement of the arbitration implementation exactly matches the design purpose of the existing systems listed in Section \ref{sec:sec:exist}, thus different mechanisms can be applied here interchangeably. 

During the operation, the user first registers their workload task. The miners register their serving machines with GPUs and stake at the arbitration module. Upon finding matching resource for the registered tasks, the serving pipeline is built on the miner-provided GPUs and the signed results are returned to the user. The corresponding payment is also processed and transferred to the miner account.
The user may monitor the received results and optionally send them to third parties for further correctness verification. If suspicious incorrect results are found, the user invokes the arbitration procedure.
With such a design, we follow the above principles that the only extra serving computation cost for the miner is to provide a signature, all honest requests remain off-chain, and the peer-to-peer communication for each task is always between the same pair of peers.

\subsection{Open Problems}

While the \sysname framework design addresses challenges in offline serving algorithms and integrates mechanisms for correctness protection, several open challenges remain. For instance, the responsibility for deploying and maintaining on-chain component smart contracts, as well as matching compute resources with tasks, remains centralized. Additionally, fair task distribution strategies for miners have yet to be fully explored. Furthermore, criteria for verifying result correctness—an area challenging for most correctness protection mechanisms—are not yet well-defined. Future research into these issues would greatly enhance the practicality of decentralized serving solutions.

\section{Related Work}

Recent work has focused extensively on enhancing GPU utilization efficiency. Some approaches prioritize computation reorganization~\cite{agrawal_taming_2024,cheng_slice-level_2024,hu_inference_2024,strati_dejavu_2024,he_deferred_2024} at GPU usage level, as outlined in a comprehensive survey~\cite{li_llm_2024}, while others optimize at the cluster level by refining device selection and scheduling for inference requests, targeting both edge~\cite{yang_perllm_2024,zhang_edgeshard_2024} and decentralized environments~\cite{mei_helix_2024}. 
In this work, we specifically target GPU usage level optimizations tailored for decentralized environments.

There is also work on the decentralized use of consumer-grade GPUs for LLMs, where some focus on training~\cite{tang_fusionai_2023,tang_fusionllm_2024}, and others on network construction~\cite{olshansky_decentralized_2024}. However, these efforts do not address the specific technical challenges associated with orchestrating LLMs over high-latency networks. Additional work has explored communication compression and model quantization to optimize performance~\cite{borzunov_petals_2023}, which affects the model behavior, and thus is orthogonal to our approach.

\bibliography{example_paper,zotero}

\begin{thebibliography}{45}
\providecommand{\natexlab}[1]{#1}
\providecommand{\url}[1]{\texttt{#1}}
\expandafter\ifx\csname urlstyle\endcsname\relax
  \providecommand{\doi}[1]{doi: #1}\else
  \providecommand{\doi}{doi: \begingroup \urlstyle{rm}\Url}\fi

\bibitem[Agrawal et~al.(2024)Agrawal, Kedia, Panwar, Mohan, Kwatra, Gulavani, Tumanov, and Ramjee]{agrawal_taming_2024}
Agrawal, A., Kedia, N., Panwar, A., Mohan, J., Kwatra, N., Gulavani, B., Tumanov, A., and Ramjee, R.
\newblock Taming \{{Throughput}-{Latency}\} {Tradeoff} in \{{LLM}\} {Inference} with \{{Sarathi}-{Serve}\}.
\newblock pp.\  117--134, 2024.
\newblock ISBN 978-1-939133-40-3.
\newblock URL \url{https://www.usenix.org/conference/osdi24/presentation/agrawal}.

\bibitem[Anthropic()]{anthropicprice}
Anthropic.
\newblock Pricing \ anthropic.
\newblock URL \url{https://www.anthropic.com/pricing#anthropic-api}.
\newblock Accessed: 2024-10-31.

\bibitem[AWS({\natexlab{a}})]{aws}
AWS.
\newblock Amazon web services (aws), {\natexlab{a}}.
\newblock URL \url{https://aws.amazon.com/}.
\newblock Accessed: 2024-10-31.

\bibitem[AWS({\natexlab{b}})]{bedrockprice}
AWS.
\newblock Amazon bedrock pricing, {\natexlab{b}}.
\newblock URL \url{https://aws.amazon.com/bedrock/pricing/}.
\newblock Accessed: 2024-10-31.

\bibitem[Azure()]{azure}
Azure.
\newblock Microsoft azure: Cloud computing services.
\newblock URL \url{https://azure.microsoft.com/en-us}.
\newblock Accessed: 2024-10-31.

\bibitem[Borzunov et~al.(2023)Borzunov, Baranchuk, Dettmers, Ryabinin, Belkada, Chumachenko, Samygin, and Raffel]{borzunov_petals_2023}
Borzunov, A., Baranchuk, D., Dettmers, T., Ryabinin, M., Belkada, Y., Chumachenko, A., Samygin, P., and Raffel, C.
\newblock Petals: {Collaborative} {Inference} and {Fine}-tuning of {Large} {Models}, March 2023.
\newblock URL \url{http://arxiv.org/abs/2209.01188}.
\newblock arXiv:2209.01188.

\bibitem[Cheng et~al.(2024)Cheng, Hu, Wang, Peng, Li, and Zhang]{cheng_slice-level_2024}
Cheng, K., Hu, W., Wang, Z., Peng, H., Li, J., and Zhang, S.
\newblock Slice-{Level} {Scheduling} for {High} {Throughput} and {Load} {Balanced} {LLM} {Serving}, June 2024.
\newblock URL \url{http://arxiv.org/abs/2406.13511}.
\newblock arXiv:2406.13511.

\bibitem[Chiang et~al.(2024)Chiang, Zheng, Sheng, Angelopoulos, Li, Li, Zhang, Zhu, Jordan, Gonzalez, et~al.]{chiang2024chatbot}
Chiang, W.-L., Zheng, L., Sheng, Y., Angelopoulos, A.~N., Li, T., Li, D., Zhang, H., Zhu, B., Jordan, M., Gonzalez, J.~E., et~al.
\newblock Chatbot arena: An open platform for evaluating llms by human preference.
\newblock \emph{arXiv preprint arXiv:2403.04132}, 2024.

\bibitem[Cloud()]{gcp}
Cloud, G.
\newblock Google cloud: Cloud computing services.
\newblock URL \url{https://cloud.google.com/}.
\newblock Accessed: 2024-10-31.

\bibitem[Conway et~al.(2024)Conway, So, Yu, and Wong]{conway2024opml}
Conway, K., So, C., Yu, X., and Wong, K.
\newblock opml: Optimistic machine learning on blockchain.
\newblock \emph{arXiv preprint arXiv:2401.17555}, 2024.

\bibitem[Dubey et~al.(2024)Dubey, Jauhri, Pandey, Kadian, Al-Dahle, Letman, Mathur, Schelten, Yang, Fan, et~al.]{dubey2024llama}
Dubey, A., Jauhri, A., Pandey, A., Kadian, A., Al-Dahle, A., Letman, A., Mathur, A., Schelten, A., Yang, A., Fan, A., et~al.
\newblock The llama 3 herd of models.
\newblock \emph{arXiv preprint arXiv:2407.21783}, 2024.

\bibitem[GIMPS()]{GIMPS}
GIMPS.
\newblock Great internet mersenne prime search.
\newblock URL \url{https://www.mersenne.org/}.
\newblock Accessed: 2024-10-31.

\bibitem[Graves(2013)]{graves2013generating}
Graves, A.
\newblock Generating sequences with recurrent neural networks.
\newblock \emph{arXiv preprint arXiv:1308.0850}, 2013.

\bibitem[He et~al.(2024)He, Lu, and Alonso]{he_deferred_2024}
He, Y., Lu, Y., and Alonso, G.
\newblock Deferred {Continuous} {Batching} in {Resource}-{Efficient} {Large} {Language} {Model} {Serving}.
\newblock In \emph{Proceedings of the 4th {Workshop} on {Machine} {Learning} and {Systems}}, pp.\  98--106, Athens Greece, April 2024. ACM.
\newblock ISBN 9798400705410.
\newblock \doi{10.1145/3642970.3655835}.
\newblock URL \url{https://dl.acm.org/doi/10.1145/3642970.3655835}.

\bibitem[Hu et~al.(2024)Hu, Huang, Xu, Chen, Xu, Chen, Feng, Wang, Wang, Bao, Sun, and Shan]{hu_inference_2024}
Hu, C., Huang, H., Xu, L., Chen, X., Xu, J., Chen, S., Feng, H., Wang, C., Wang, S., Bao, Y., Sun, N., and Shan, Y.
\newblock Inference without {Interference}: {Disaggregate} {LLM} {Inference} for {Mixed} {Downstream} {Workloads}, January 2024.
\newblock URL \url{http://arxiv.org/abs/2401.11181}.
\newblock arXiv:2401.11181.

\bibitem[Huang et~al.(2019)Huang, Cheng, Bapna, Firat, Chen, Chen, Lee, Ngiam, Le, Wu, et~al.]{huang2019gpipe}
Huang, Y., Cheng, Y., Bapna, A., Firat, O., Chen, D., Chen, M., Lee, H., Ngiam, J., Le, Q.~V., Wu, Y., et~al.
\newblock Gpipe: Efficient training of giant neural networks using pipeline parallelism.
\newblock \emph{Advances in neural information processing systems}, 32, 2019.

\bibitem[io.net()]{io.net}
io.net.
\newblock io.net.
\newblock URL \url{https://io.net/}.
\newblock Accessed: 2024-10-31.

\bibitem[Kalodner et~al.(2018)Kalodner, Goldfeder, Chen, Weinberg, and Felten]{kalodner2018arbitrum}
Kalodner, H., Goldfeder, S., Chen, X., Weinberg, S.~M., and Felten, E.~W.
\newblock Arbitrum: Scalable, private smart contracts.
\newblock In \emph{27th USENIX Security Symposium (USENIX Security 18)}, pp.\  1353--1370, 2018.

\bibitem[Li et~al.(2024)Li, Jiang, Gadepally, and Tiwari]{li_llm_2024}
Li, B., Jiang, Y., Gadepally, V., and Tiwari, D.
\newblock {LLM} {Inference} {Serving}: {Survey} of {Recent} {Advances} and {Opportunities}, July 2024.
\newblock URL \url{http://arxiv.org/abs/2407.12391}.
\newblock arXiv:2407.12391.

\bibitem[Li et~al.(2021)Li, Zhuang, Guo, Zhuo, Zhang, Song, and Stoica]{li2021terapipe}
Li, Z., Zhuang, S., Guo, S., Zhuo, D., Zhang, H., Song, D., and Stoica, I.
\newblock Terapipe: Token-level pipeline parallelism for training large-scale language models.
\newblock In \emph{International Conference on Machine Learning}, pp.\  6543--6552. PMLR, 2021.

\bibitem[Li et~al.(2023)Li, Zheng, Zhong, Liu, Sheng, Jin, Huang, Chen, Zhang, Gonzalez, et~al.]{li2023alpaserve}
Li, Z., Zheng, L., Zhong, Y., Liu, V., Sheng, Y., Jin, X., Huang, Y., Chen, Z., Zhang, H., Gonzalez, J.~E., et~al.
\newblock $\{$AlpaServe$\}$: Statistical multiplexing with model parallelism for deep learning serving.
\newblock In \emph{17th USENIX Symposium on Operating Systems Design and Implementation (OSDI 23)}, pp.\  663--679, 2023.

\bibitem[Marr(2023)]{companyUseGPT}
Marr, B.
\newblock 10 amazing real-world examples of how companies are using chatgpt in 2023, 2023.
\newblock URL \url{https://www.forbes.com/sites/bernardmarr/2023/05/30/10-amazing-real-world-examples-of-how-companies-are-using-chatgpt-in-2023}.

\bibitem[Mei et~al.(2024)Mei, Zhuang, Miao, Yang, Jia, and Vinayak]{mei_helix_2024}
Mei, Y., Zhuang, Y., Miao, X., Yang, J., Jia, Z., and Vinayak, R.
\newblock Helix: {Distributed} {Serving} of {Large} {Language} {Models} via {Max}-{Flow} on {Heterogeneous} {GPUs}, June 2024.
\newblock URL \url{http://arxiv.org/abs/2406.01566}.
\newblock arXiv:2406.01566.

\bibitem[Nakamoto(2008)]{nakamoto2008bitcoin}
Nakamoto, S.
\newblock Bitcoin: A peer-to-peer electronic cash system.
\newblock \emph{Satoshi Nakamoto}, 2008.

\bibitem[Olshansky et~al.(2024)Olshansky, Colmeiro, and Li]{olshansky_decentralized_2024}
Olshansky, D., Colmeiro, R.~R., and Li, B.
\newblock Decentralized {AI}: {Permissionless} {LLM} {Inference} on {POKT} {Network}, May 2024.
\newblock URL \url{http://arxiv.org/abs/2405.20450}.
\newblock arXiv:2405.20450.

\bibitem[OpenAI({\natexlab{a}})]{openaibatch}
OpenAI.
\newblock Batch - openai api, {\natexlab{a}}.
\newblock URL \url{https://platform.openai.com/docs/guides/batch}.
\newblock Accessed: 2024-10-31.

\bibitem[OpenAI({\natexlab{b}})]{openaiprice}
OpenAI.
\newblock Pricing | openai, {\natexlab{b}}.
\newblock URL \url{https://openai.com/api/pricing/}.
\newblock Accessed: 2024-10-31.

\bibitem[Orth(2023)]{yougov}
Orth, T.
\newblock What americans think about chatgpt and ai-generated text, 2023.
\newblock URL \url{https://today.yougov.com/technology/articles/45128-what-americans-think-about-chatgpt-and-ai-text}.

\bibitem[Paszke et~al.(2019)Paszke, Gross, Massa, Lerer, Bradbury, Chanan, Killeen, Lin, Gimelshein, Antiga, Desmaison, Kopf, Yang, DeVito, Raison, Tejani, Chilamkurthy, Steiner, Fang, Bai, and Chintala]{torch}
Paszke, A., Gross, S., Massa, F., Lerer, A., Bradbury, J., Chanan, G., Killeen, T., Lin, Z., Gimelshein, N., Antiga, L., Desmaison, A., Kopf, A., Yang, E., DeVito, Z., Raison, M., Tejani, A., Chilamkurthy, S., Steiner, B., Fang, L., Bai, J., and Chintala, S.
\newblock Pytorch: An imperative style, high-performance deep learning library.
\newblock In \emph{Advances in Neural Information Processing Systems 32}, pp.\  8024--8035. Curran Associates, Inc., 2019.
\newblock URL \url{http://papers.neurips.cc/paper/9015-pytorch-an-imperative-style-high-performance-deep-learning-library.pdf}.

\bibitem[Pope et~al.(2023)Pope, Douglas, Chowdhery, Devlin, Bradbury, Heek, Xiao, Agrawal, and Dean]{pope2023efficiently}
Pope, R., Douglas, S., Chowdhery, A., Devlin, J., Bradbury, J., Heek, J., Xiao, K., Agrawal, S., and Dean, J.
\newblock Efficiently scaling transformer inference.
\newblock \emph{Proceedings of Machine Learning and Systems}, 5:\penalty0 606--624, 2023.

\bibitem[RunPod()]{runpod}
RunPod.
\newblock Runpod - the cloud built for ai.
\newblock URL \url{https://www.runpod.io/}.
\newblock Accessed: 2024-10-31.

\bibitem[Shazeer et~al.(2018)Shazeer, Cheng, Parmar, Tran, Vaswani, Koanantakool, Hawkins, Lee, Hong, Young, et~al.]{shazeer2018mesh}
Shazeer, N., Cheng, Y., Parmar, N., Tran, D., Vaswani, A., Koanantakool, P., Hawkins, P., Lee, H., Hong, M., Young, C., et~al.
\newblock Mesh-tensorflow: Deep learning for supercomputers.
\newblock \emph{Advances in neural information processing systems}, 31, 2018.

\bibitem[Shoeybi et~al.(2019)Shoeybi, Patwary, Puri, LeGresley, Casper, and Catanzaro]{shoeybi2019megatron}
Shoeybi, M., Patwary, M., Puri, R., LeGresley, P., Casper, J., and Catanzaro, B.
\newblock Megatron-lm: Training multi-billion parameter language models using model parallelism.
\newblock \emph{arXiv preprint arXiv:1909.08053}, 2019.

\bibitem[Strati et~al.(2024)Strati, Mcallister, Phanishayee, Tarnawski, and Klimovic]{strati_dejavu_2024}
Strati, F., Mcallister, S., Phanishayee, A., Tarnawski, J., and Klimovic, A.
\newblock D{éjàVu}: {KV}-cache {Streaming} for {Fast}, {Fault}-tolerant {Generative} {LLM} {Serving}, March 2024.
\newblock URL \url{http://arxiv.org/abs/2403.01876}.
\newblock arXiv:2403.01876.

\bibitem[Sun et~al.(2024)Sun, Li, and Zhang]{sun2024zkllm}
Sun, H., Li, J., and Zhang, H.
\newblock zkllm: Zero knowledge proofs for large language models.
\newblock \emph{arXiv preprint arXiv:2404.16109}, 2024.

\bibitem[Tang et~al.(2023)Tang, Wang, He, Zhang, Pan, Wang, Zeng, Zhao, Shi, He, and Chu]{tang_fusionai_2023}
Tang, Z., Wang, Y., He, X., Zhang, L., Pan, X., Wang, Q., Zeng, R., Zhao, K., Shi, S., He, B., and Chu, X.
\newblock {FusionAI}: {Decentralized} {Training} and {Deploying} {LLMs} with {Massive} {Consumer}-{Level} {GPUs}, September 2023.
\newblock URL \url{http://arxiv.org/abs/2309.01172}.
\newblock arXiv:2309.01172.

\bibitem[Tang et~al.(2024)Tang, Kang, Yin, Pan, Wang, He, Wang, Zeng, Zhao, Shi, Zhou, Li, He, and Chu]{tang_fusionllm_2024}
Tang, Z., Kang, X., Yin, Y., Pan, X., Wang, Y., He, X., Wang, Q., Zeng, R., Zhao, K., Shi, S., Zhou, A.~C., Li, B., He, B., and Chu, X.
\newblock {FusionLLM}: {A} {Decentralized} {LLM} {Training} {System} on {Geo}-distributed {GPUs} with {Adaptive} {Compression}, October 2024.
\newblock URL \url{http://arxiv.org/abs/2410.12707}.
\newblock arXiv:2410.12707.

\bibitem[WhatToMine()]{whattomine}
WhatToMine.
\newblock Whattomine.
\newblock URL \url{https://whattomine.com/}.
\newblock Accessed: 2024-10-31.

\bibitem[Yang et~al.(2024)Yang, Yang, Zhao, Guo, He, and Ji]{yang_perllm_2024}
Yang, Z., Yang, Y., Zhao, C., Guo, Q., He, W., and Ji, W.
\newblock {PerLLM}: {Personalized} {Inference} {Scheduling} with {Edge}-{Cloud} {Collaboration} for {Diverse} {LLM} {Services}, May 2024.
\newblock URL \url{http://arxiv.org/abs/2405.14636}.
\newblock arXiv:2405.14636.

\bibitem[Ye et~al.()Ye, Chen, Lai, Zhao, Zheng, Shao, Hou, Jin, Zuo, Yin, Chen, and Ceze]{flashinfer2024}
Ye, Z., Chen, L., Lai, R., Zhao, Y., Zheng, S., Shao, J., Hou, B., Jin, H., Zuo, Y., Yin, L., Chen, T., and Ceze, L.
\newblock Accelerating {{Self-Attentions}} for {{LLM Serving}} with {{FlashInfer}}.
\newblock URL \url{https://flashinfer.ai/2024/02/02/introduce-flashinfer.html}.

\bibitem[Yu et~al.(2022)Yu, Jeong, Kim, Kim, and Chun]{yu2022orca}
Yu, G.-I., Jeong, J.~S., Kim, G.-W., Kim, S., and Chun, B.-G.
\newblock Orca: A distributed serving system for $\{$Transformer-Based$\}$ generative models.
\newblock In \emph{16th USENIX Symposium on Operating Systems Design and Implementation (OSDI 22)}, pp.\  521--538, 2022.

\bibitem[Zhang et~al.(2024{\natexlab{a}})Zhang, Cao, Shen, and Cui]{zhang_edgeshard_2024}
Zhang, M., Cao, J., Shen, X., and Cui, Z.
\newblock {EdgeShard}: {Efficient} {LLM} {Inference} via {Collaborative} {Edge} {Computing}, May 2024{\natexlab{a}}.
\newblock URL \url{http://arxiv.org/abs/2405.14371}.
\newblock arXiv:2405.14371.

\bibitem[Zhang et~al.(2024{\natexlab{b}})Zhang, Wang, Liu, Tan, Popa, and Moallemi]{zhang2024proof}
Zhang, Y., Wang, S., Liu, X., Tan, S., Popa, R.~A., and Moallemi, C.~C.
\newblock Proof of sampling: A nash equilibrium-secured verification protocol for decentralized systems.
\newblock \emph{arXiv preprint arXiv:2405.00295}, 2024{\natexlab{b}}.

\bibitem[Zheng et~al.(2022)Zheng, Li, Zhang, Zhuang, Chen, Huang, Wang, Xu, Zhuo, Xing, Gonzalez, and Stoica]{zheng2022alpa}
Zheng, L., Li, Z., Zhang, H., Zhuang, Y., Chen, Z., Huang, Y., Wang, Y., Xu, Y., Zhuo, D., Xing, E.~P., Gonzalez, J.~E., and Stoica, I.
\newblock Alpa: Automating inter- and intra-operator parallelism for distributed deep learning, 2022.

\bibitem[Zheng et~al.(2021)Zheng, Xie, Zhang, Chen, Chen, Guo, Sun, Sun, and Zhou]{zheng2021agatha}
Zheng, Z., Xie, P., Zhang, X., Chen, S., Chen, Y., Guo, X., Sun, G., Sun, G., and Zhou, L.
\newblock Agatha: Smart contract for dnn computation.
\newblock \emph{arXiv preprint arXiv:2105.04919}, 2021.

\end{thebibliography}
\bibliographystyle{mlsys2025}

% \appendix
% \input{content/appendix}

\end{document}